\documentclass[conference]{IEEEtran}
\IEEEoverridecommandlockouts
\usepackage{cite}
\usepackage{amsmath,amssymb,amsfonts}
\usepackage{algorithmic}
\pdfoutput=1
\usepackage{graphicx}
\usepackage{textcomp}
\usepackage{xcolor}
\usepackage{times}
\usepackage{fancyhdr,graphicx,amsmath,amssymb}
\usepackage{floatrow}  
\include{pythonlisting}
\usepackage{flushend}
\usepackage{float}   
\usepackage{makecell}
\usepackage{multirow}
\usepackage{booktabs}
\usepackage[caption=false,font=footnotesize,labelfont=rm,textfont=rm]{subfig}
\usepackage{hyperref}
\hypersetup{hidelinks,
	colorlinks=true,
	allcolors=black,
	pdfstartview=Fit,
	breaklinks=true}
\usepackage{url}

\def\BibTeX{{\rm B\kern-.05em{\sc i\kern-.025em b}\kern-.08em
    T\kern-.1667em\lower.7ex\hbox{E}\kern-.125emX}}

\begin{document}

\title{Resilient Random Time-hopping Reply against Distance Attacks in UWB Ranging
\thanks{This work was supported in part by the Fundamental Research Funds for the Central Universities under Grant 2242022k60006. }
}

\author{
\IEEEauthorblockN{Wenlong Gou, Chuanhang Yu, Gang Wu}
\IEEEauthorblockA{National Key Laboratory of Wireless Communications,\\
University of Electronic Science and Technology of China, Chengdu, China}

{\{gouwenlong, chuanhangyu\}}@std.uestc.edu.cn,\\ wugang99@uestc.edu.cn (corresponding author)
}

\maketitle

\begin{abstract}
In order to mitigate the distance reduction attack in Ultra-Wide Band (UWB) ranging, this paper proposes a secure ranging scheme based on a random time-hopping mechanism without redundant signaling overhead, which is also backward
compatible with existing standards such as IEEE 802.15.4a/z.
By randomly varying the reply interval between ranging messages while maintaining synchronization of the interval between legitimate devices, distance attacks can be effectively defended against.
The effectiveness and feasibility of the proposed strategy are demonstrated through simulation and experiments in the case of the \emph{Ghost Peak} attack, which can effectively reduce the ranging distance of existing commercial UWB chips. 
The random time-hopping mechanism is verified to be capable of reducing the success rate of distance reduction attacks to less than 0.01{\%}, thereby significantly enhancing the security of UWB ranging.

\end{abstract}

\begin{IEEEkeywords}
Ultra-Wideband, secure ranging, distance reduction attack, random
time-hopping mechanism
\end{IEEEkeywords}

\section{Introduction}
With non-sine-wave narrow pulses for ranging, positioning and data transmission, Ultra-Wideband (UWB) has the characteristics of strong anti-multipath fading ability and low power consumption, and can provide centimeter-level ranging precision, making it widely applicable in ranging and short-range communication scenarios{\cite{y1}}{\cite{y2}}. 
However, there always exist numerous security threats in UWB ranging. For instance, the IEEE 802.15.4a protocol is vulnerable to the \emph{Cicada} attack  which continuously injects uniformly-spaced UWB pulses into the receiver during the legitimate transmission of the preamble \cite{y3}, and the Early Detection/Late Commitment (ED/LC) attack which leverages the predictability of the signal structure within the preamble{\cite{y4}}.
To address known range-reduction attacks, IEEE has released a new version standard 802.15.4z{\cite{y5}}, which enhances the ranging precision and security by introducing Scrambled Timestamp Sequences (STS) encrypted by the Advanced Encryption Standard (AES). 
But in {\cite{y6}}, Miridula Singh proposed the \emph{Cicada++} attack which executes the distance attack by transmitting pseudo-random STS signals with a fixed pulse repetition frequency to alter the timestamps of received signals, and further proposed the Adaptive Injection Attack (AIA) which can refine the attack precision by controlling the placement of injected attack signals. 
Patrick Leu et al.{\cite{y7}} demonstrated the \emph{Ghost Peak} attack achieving a success rate of up to 4\% on commercially available Apple U1 and Qorvo UWB chips. In {\cite{y8}}, Claudio Anliker et al. also proposed the \emph{Mix-Down} attack, which exploits the clock drift of transceivers. 
Recently, an effective jamming attack by sniffing the fixed time interval between two ranging packets was successfully conducted in {\cite{v1}}.     
These studies confirm the persistence of security gaps in the new standard.

Several methods have been proposed to defend against distance attacks. For example , a ranging scheme combining Time of Flight (TOF) and Received Signal Strength (RSS) is proposed in {\cite{y9}} to effectively mitigate distance fraud. 
Additionally, the authors in {\cite{y10}} designed the UnSpoof UWB localization system, capable of pinpointing the position of both the attacker and the legitimate device. 
Meanwhile, effectively detecting attacks in the process of UWB security ranging is also a problem to be considered. 
In {\cite{y12}}, Mridula Singh presented a novel modulation technique for detecting distance enlargement attacks based on the interleaving of pulses of different phases. 
The scheme in{\cite{y13}} achieved a 96.24\% success rate in attack detection by leveraging the consistency of cross-correlation results between the sub-fields of STS and local templates.

Different from most of the defense schemes against distance
attacks {\cite{y9,y10,y12}} by modifying the established UWB physical layer standards, a security ranging scheme adopting a random time-hopping mechanism is proposed, which integrates an attack detection scheme using channel reciprocity and autoencoder {\cite{y15}}, can decrease the costs of practical deployments and signaling overhead.
Changing the synchronization time between the attack and legitimate signals prevents distance reduction attacks without altering established standards.
The scheme is effective against attacks that require prior knowledge by sniffing legitimate message transmissions and receptions, such as \emph{Cicada++}, \emph{Ghost Peak}, ED/LC, Adaptive Injection, etc. And the superior performance of this strategy under the \emph{Ghost Peak} attack is verified in this paper.
The source code is available at \url{https://github.com/cn-wlg/UWB_source}. 
Our main contributions can be summarized as follows:
\begin{figure}[htbp]
               
			\centering   
			
			\includegraphics[width=1\linewidth]{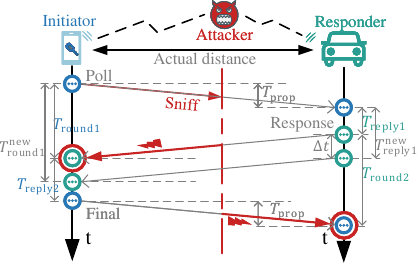}
			\caption{Classic UWB ranging model and random time-hopping ranging model}
                \label{fign1}
                 \vspace{-2mm}
\end{figure}
\begin{itemize}
    \item A security ranging scheme against distance attacks based on random time-hopping is proposed, along with its theoretical derivation. The effectiveness of the scheme is verified by simulation, and the feasibility is validated on a commercial hardware platform.
    \item An STS-aided synchronization method is proposed to ensure that both the transmitter and receiver are synchronized regarding the time-hopping value  without consuming redundant signaling. By pre-storing the optional time-hopping value as a hash table and then randomly selecting it, the difficulty of subsequent attacks is increased dramatically.
\end{itemize}

The rest of the paper is organized as follows. Section \ref{System Model} describes system model and problem statement. The designed security ranging scheme is detailed in Section \ref{Security Ranging Strategy Design}. Section \ref{ Simulation Evaluation and Experimental Validation} presents evaluation. Finally, Section \ref{Conclusion} concludes this paper.

\section{System Model and problem Statement}
\label{System Model}

\subsection{Classic UWB Ranging Model and Attack}

To mitigate the effects of factors such as clock drift, Double Side-Two Way Ranging (DS-TWR) is commonly used in UWB systems, which requires three ranging messages to be exchanged between both ranging sides. 
The distance is estimated as{\cite{y5}}:
\begin{equation}
\label{eq1}
  d=c\cdot T_{\text{prop}}=\frac{T_{\text{{round}}_\text{1}}\times T_{\text{{round}}_\text{2}}-T_{\text{{ reply}}_\text{1}}\times T_{\text{{ reply}}_\text{2}}}{T_{\text{{ round}}_\text{1}}+T_{\text{{ round}}_\text{2}}+T_{\text{{reply}}_\text{1}}+T_{\text{{ reply}}_\text{2}}}c,  
\end{equation}
where the time intervals $T_{\text{{round}}_\text{1}}$, $T_{\text{{round}}_\text{2}}$, $T_{\text{{ reply}}_\text{1}}$, $T_{\text{{ reply}}_\text{2}}$ are shown in Fig.~\ref{fign1}, $c$ refers to the speed of light.

\begin{figure}[tbp]
               
			\centering 
                \subfloat[{Packet Configuration 1 in IEEE 802.15.4z{\cite{y5}}}]{
			\includegraphics[width=0.85\linewidth]{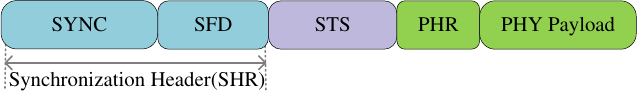}
                \label{figpacket}
                }
                \newline                
                
			\subfloat[Cross-correlation spectrum between received STS and loacl STS]{
			\includegraphics[width=0.85\linewidth]{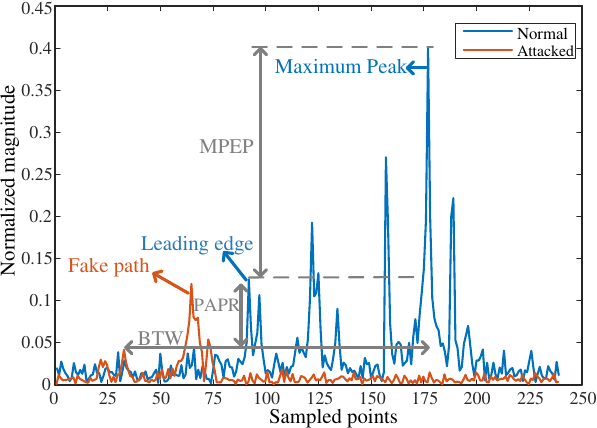}
                \label{fign2}
                }   
                
                 \caption{Tmestamp acquisition}
                 \vspace{-2mm}
                
\end{figure}

Fig.~\ref{figpacket} illustrates a typical packet in IEEE 802.15.4z{\cite{y5}}, which consists of the Synchronization preamble (SYNC), Start Frame Delimiter (SFD), STS, Physical Header (PHR) and PHY Payload field. 
The timestamps in the DS-TWR process in High Rate Pulse (HRP) mode{\cite{y5}} are obtained using a leading edge detection algorithm, which operates on the cross-correlation spectrum of the STS segments to determine the earliest arrival time of the received signal. 
As shown in Fig.~\ref{fign2}, the Maximum Peak to Early Peak Ratio (MPEP), which represents the ratio between the main path and the first path power, and the Peak to Average Power Ratio (PAPR), which represents the ratio between the peak and the average power, are two critical thresholds of the leading edge detection algorithm based on the Back-Search Time Window (BTW). 
The further accurate estimation of timestamps by the leading edge detection algorithm also provides attackers with more opportunities to increase the attack success probability, which is mainly exploited by many distance attacks represented by the \emph{Ghost Peak} attack.

STS encrypted with AES are typically considered unpredictable, which is the premise of various distance attacks.
In the \emph{Ghost Peak} attack shown in Fig.~\ref{fign1}, after the attacker sniffs the transmit time of the ranging message, it would send the attack signal during the reception of the Response message or Final message of the legitimate device. The STS segment of the attack signal is forged and its power is much higher than that of the legitimate signal. As shown in Fig.~\ref{fign2}, the \emph{Ghost Peak} attack results in a fake path in the cross-correlation spectrum.
Additionally, the attack synchronization time, defined as the time difference between the transmit time of the attack message and the corresponding legitimate message, significantly impacts the attack success rate in the \emph{Ghost Peak} attack, and this will be elaborated in detail in the simulation results in Sec.~\ref{Numerical Simulation Evaluation}.





\subsection{Random Time-hopping Ranging Model Against Attacks}
\label{Random Time-hopping}

For distance attacks which require the transmit time of the attack signal to be highly aligned with that of the legitimate signal, we propose a secure ranging scheme based on the random time-hopping mechanism. 
Similar to the frequency-hopping technique applied in spread spectrum communication systems with robust anti-interference capability, the random time-hopping mechanism ensures the security of UWB ranging by randomly transmitting messages at different time instants to increase the attack synchronization time, i.e., reducing the degree of alignment between the attack signal and the corresponding legitimate signal transmit time.
 \begin{figure*}[htbp]              
    \centering   
    \centerline{\includegraphics[width=1\linewidth]{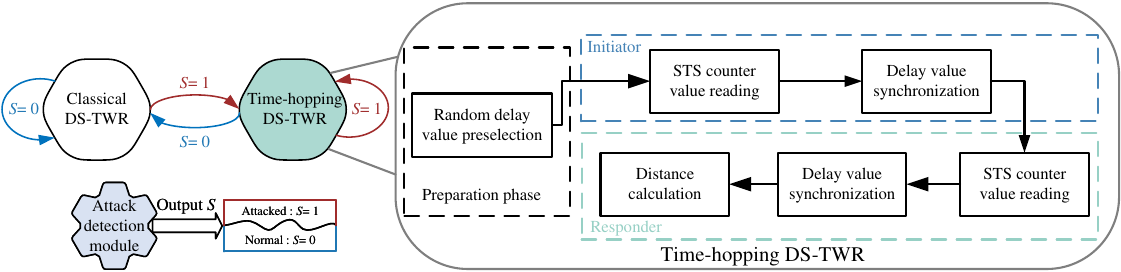}}
    \caption{A security ranging strategy with attack detection and time-hopping DS-TWR for UWB}
    \label{fign3}
    
    \vspace{-2mm}

\end{figure*}

Taking the \emph{Ghost Peak} attack on the Response message as an example, Fig.~\ref{fign1} demonstrates the anti-attack ranging model based on random time-hopping. 
The Responder selects a delay $\Delta t $ that can resist the attack without disrupting the overall ranging process, so that the Response message is delayed with  $\Delta t $ to be transmitted. The modified $T_{\text{{round}}_\text{1}}^{\text{new}}$ and $T_{\text{{reply}}_\text{1}}^{\text{new}}$ are updated based on the selected $\Delta t $ and then continue to complete the ranging process. 
This approach results in a large difference between the arrival time of the legitimate Response message and the attack signal at the Initiator, thus avoiding the occurrence of the distance variation caused by the attack.

\subsection{Problem Statement}

In the \emph{Ghost Peak} attack, the timestamp estimated by the leading edge detection algorithm at the time offset $t$ is valid if the forged STS segment used by the attacker satisfies:
\begin{small}
\begin{equation}
\label{eq2}
\left| {x_t} {\sum\limits_{i = 1}^{\text{N}} \int_{(i-1)T_\text{b}}^{iT_\text{b}} {a_{\text{L}}\lbrack i\rbrack p(\tau) a_{\text{A}}\lbrack i - n\rbrack p^{*}(\tau)}d\tau}\right|>\max\left\{ P_{\text{max}}\text{T}_{\text{m}},P_{\text{rms}}\text{T}_{\text{p}} \right\} 
\end{equation}
\end{small}where $t=\frac{n}{\text{F}_\text{{s}}}\left( n = 1,2,\cdots,\text{T}_\text{{BTW}}\text{F}_\text{{s}} \right) $ denotes  the time corresponding to the sampling point, $\text{F}_\text{{s}}$ is the sampling rate, $\text{T}_\text{{BTW}} $ is the size of the BTW, $T_\text{b}$ is the duration of a symbol, $P_{\text{max}}$ and $ P_{\text{rms}}$ are the maximum and root mean square power of the cross-correlation spectrum, $\text{T}_{\text{m}}$ and $\text{T}_{\text{p}}$ are the thresholds MPEP and PAPR corresponding to the leading edge detection algorithm, $
a_{\text{L}}\lbrack n\rbrack,a_{\text{A}}\lbrack n\rbrack \in \left\{ - 1,1 \right\} $ are binary STS codes of the local template and attacker, $p(\tau)$ is the pulse shape with unit power modulated by Binary Phase-Shift Keying (BPSK)\cite{y5}, $\text{N}$ is the length of STS, and ${x_t}$ represents the adversarial signal power at offset $t$ \cite{g1}.

Since STS is a pseudo-random sequence, it should satisfy:
\begin{equation} 
\label{eq3}
{\mathbb{P}({a_{\text{L}}\lbrack i\rbrack  a_{\text{A}}\lbrack i - n\rbrack= - 1})}= {\mathbb{P}({a_{\text{L}}\lbrack i\rbrack  a_{\text{A}}\lbrack i - n\rbrack= 1})} = 0.5.
\end{equation}

Let $ X \backsim \mathcal{B} (\text{N},0.5)$ be a random variable following binomial distribution, then the random variable $ \sum\limits_{i = 0}^{\text{N}}{a_{\text{L}}\lbrack i\rbrack~a_{\text{A}}\lbrack i - n\rbrack}$ follows:
\begin{equation}
\label{eq4}
    {\sum\limits_{i = 0}^{\text{N}}{a_{\text{L}}\lbrack i\rbrack~a_{\text{A}}\lbrack i - n\rbrack}} = 2X - \text{N}.
\end{equation}

For simplicity, let $ \max\left\{ P_{\text{max}}\text{T}_{\text{m}},~P_{\text{rms}}\text{T}_{\text{p}} \right\} = \theta$, thus the attack success probability $ {\mathbb{P}}_{\text{s}}(t) $ at the time offset $t$ is: 
\begin{equation}
\label{eq5}
\begin{aligned}
   {\mathbb{P}_\text{s}}(t) &=\mathbb{P}\left( {\left| {2 X - \text{N}} \right|\left| {p(t)} \right|^{2}{x_t}  > \theta} \right)\\ 
   &=2\mathbb{P}\left( { X > 0.5\frac{ \theta}{x_t}}+0.5\text{N} \right)\\
   &\leq 2exp\left( {- \frac{\theta ^{2}}{2{x_t}^{2}\text{N}} }\right),
\end{aligned}
\end{equation}
where the upper bound is derived by Hoeffding's inequality \cite{wg1}. 
It can be concluded that the adversarial signal power exerts a significant influence on the attack success rate in addition to the threshold. This also explains why the power of the STS segment is significantly increased in the \emph{Ghost Peak} attack.

Meanwhile, a UWB receiver typically utilizes the cross-correlation
result of the STS segment as the Channel Impulse Response
(CIR) for data demodulation. 
In order to ensure correct demodulation of the Payload field without interfering with the detection of the SFD field, the time offset $t$ should satisfy the following condition:
\begin{equation}
\label{eq6}
    {T_{\text{SFD}} \leq t \leq T_{\text{payload}}}
\end{equation}
where $T_{\text{SFD}}$ and $T_{\text{payload}} $ refer to the thresholds greater than the end instant of the SFD field and less than the beginning instant of the Payload field, respectively. 
These two thresholds ensure that the receiver is able to complete the reception of the ranging message, a prerequisite for the success of the attack.
We aim to reduce the attack success rate and enhance the security of UWB ranging by breaking the requirement for synchronous alignment between the transmission times of the attack and legitimate messages.



\section{Security Ranging scheme Design}
\label{Security Ranging Strategy Design}


\subsection{Overall Procedure }

The ranging algorithm's simplicity and efficiency are vital for Internet of Things (IoT) applications. We propose a UWB secure ranging approach with a random time-hopping mechanism, comprising an attack detection module and a time-hopping DS-TWR, as depicted in Fig.~\ref{fign3}.

The purpose of attack detection is to enable the UWB ranging system to recognize an ongoing attack, indicated by the output $S$ of the attack detection module being equal to 1.
Then the ranging mode switches from the classical DS-TWR with lower power consumption to the time-hopping DS-TWR with higher power consumption but higher security. 
The principle of channel reciprocity based attack detection involves compressing and quantizing the CIR feature using an autoencoder in a complete ranging process, then transmitting them using the Payload field, and finally detecting the attack by comparing the CIR feature of both ranging sides. 
It is not the focus of this paper and the detailed algorithm can be referred in{\cite{y15}}. 

As described in Sec.~\ref{Random Time-hopping}, the security of the time-hopping ranging model is achieved by randomly varying the transmit time of the ranging messages. 
In brief, the random transmit time of legitimate messages makes it difficult for an attacker to accurately transmit the forged attack signal during the reception of the corresponding legitimate message, thus disrupting the synchronization conditions required for the attacks. 

The range of selectable random time-hopping delay $ \Delta t$ is broad. 
In practical UWB applications, the TOF is typically on the order of nanosecond and the delay between each ranging message is on the order of millisecond. Basically, all types of distance attack methods must ensure that the attack signal and legitimate signal arrive at the receiver at a highly consistent time. 
Therefore, it is easy to select a set of $ \Delta t$ values that can effectively invalidate the attack without disrupting the entire ranging process.

\subsection{Random Time-hopping Ranging without Redundant Signaling Overheads}

In time-hopping DS-TWR, both ranging sides also need to obtain the specific random time-hopping delay value $ \Delta t$ in real time, resulting in additional signaling overhead. 
In the generation process of the STS with AES encryption mechanism shown in Fig.~\ref{fign5}, there exists a counter whose value is incremented each time a ranging message is transmitted or received and the counter values of both ranging sides remain consistent strictly. 
Accordingly, this paper proposes an anti-attack time-hopping ranging scheme that eliminates redundant signaling, implemented as follows: 
\subsubsection{Random delay value preselection}
Sort out the available $ \Delta t$ values that invalidate the attack  without disrupting the entire ranging process and store them in a hash table for quick access. 
\subsubsection{STS counter value reading}
\begin{figure}[htbp]              
    \centering   
    \centerline{\includegraphics[width=1\linewidth]{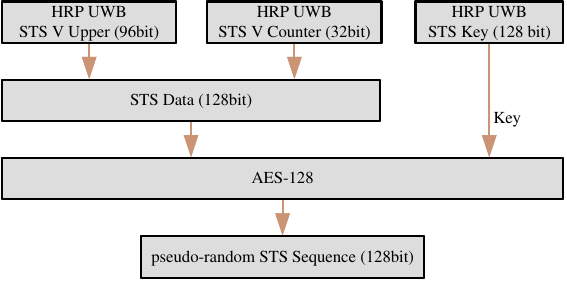}}
    \caption{STS generation procedure in IEEE 802.15.4z{\cite{y5}}}
    \label{fign5}
    \vspace{-2mm}

\end{figure}

When the Initiator and Responder transmit or receive a ranging message, read their respective counter values, i.e., 128 bits of STS Data in Fig.~\ref{fign5}. 
\subsubsection{Delay synchronization and distance calculation}
Both ranging sides use their own counter value as a random seed to randomly select a $ \Delta t$ value from the stored hash table. They then complete the ranging process using the modified $T_{\text{round1}}^{\text{new}}$ and $ T_{\text{reply1}}^{\text{new}}$.

\subsection{Performance Analysis}

It is generally assumed that $t$ follows a uniform distribution $ \text{U}(\text{T}_{\text{min}}, \text{T}_{\text{max}})$, where $\text{T}_{\text{min}}$ and $\text{T}_{\text{max}}$ represent the minimum and maximum offsets of possible TOF in practice respectively. 
Taking into consideration \eqref{eq5} and \eqref{eq6}, the attack success probability  $ {\mathbb{P}_{\text{s}}}^{'}(t)$ at the time offset $t$ becomes:

\begin{equation}
\label{eq7}
\begin{aligned}
   {\mathbb{P}_{\text{s}}}^{'}(t) &= {\mathbb{P}_\text{s}}(t)~ \mathbb{P}({T_{\text{SFD}} \leq t \leq T_{\text{payload}}})\\ 
   &=2\mathbb{P}\left( { X > 0.5\frac{ \theta}{x_t}}+0.5\text{N} \right) \frac{\Delta t_0}{\Delta t_1},
\end{aligned}
\end{equation}
where $ \Delta t_0= T_{\text{payload}} - T_{\text{SFD}} $ and $ \Delta t_1= T_{\text{max}} - T_{\text{min}} $.

If the selected random hopping delay $ \Delta t$ ranges between $T_{\text{min}}^{\text{hop}}$ and $T_{\text{max}}^{\text{hop}}$, $\Delta t$ can be considered to follow the uniform distribution $ \text{U}(T_{\text{min}}^{\text{hop}}, T_{\text{max}}^{\text{hop}})$ and be independent from $t$. Furthermore, to guarantee the success of the attack, $ \Delta t$ and $t$ should satisfy:
\begin{equation}
    \label{eq8}
    T_{\text{SFD}} \leq t - \Delta t \leq T_{\text{payload}}.
\end{equation}

Let the random variable $ Y=t - \Delta t$. The attack success probability $ {\mathbb{P}_{\text{s}}}^{''}(t)$ after adopting the random time-hopping mechanism becomes:  
\begin{equation}
\label{eq9}
\begin{aligned}
   {\mathbb{P}_{\text{s}}}^{''}(t) &= {\mathbb{P}_\text{s}}(t)~ \mathbb{P}({T_{\text{SFD}} \leq t- \Delta t \leq T_{\text{payload}}})\\ 
   &=2\mathbb{P}\left( { X > 0.5\frac{ \theta}{x_t}}+0.5\text{N} \right)\times{\int_{T_{\text{SFD}}}^{T_{\text{payload}}}f_{Y}}(y)dy\\
   &\leq 2exp\left( {- \frac{\theta ^{2}}{2{x_t}^{2}\text{N}} } \right) \times{\int_{T_{\text{SFD}}}^{T_{\text{payload}}}f_{Y}}(y)dy,
\end{aligned}
\end{equation}
where ${f_{Y}}(y)$  is the probability density function (pdf) of the random variable $Y$, obtained by convolving two uniform distributions.

Let $ \Delta t_2= T_{\text{max}}^{\text{hop}}- T_{\text{min}}^{\text{hop}} $, then ${f_{Y}}(y)$ is an isosceles trapezoid between $T_{\text{max}}^{\text{hop}}$ and $ T_{\text{min}}^{\text{hop}}$, with a height of $\frac{1}{\Delta t_2} $ and an upper base of $ \Delta t_2 -\Delta t_1$. 
Since the range of delay $\Delta t_2 $ for the proposed random time-hopping is much larger than the original possible TOF $ \Delta t_1$ (i.e. $\Delta t_2 \gg \Delta t_1$), $ {f_{Y}}(y)$ can be approximated as a rectangle with bottom $ \Delta t_2$ and height $ \frac{1}{\Delta t_2}$. 
Therefore, the reduction factor $G$ of the attack success probability, representing the gain of this strategy, is:
\begin{equation}
    G= \frac{\Delta t_2}{\Delta t_1} \rightarrow\infty 
\end{equation}

On the other hand, the complexity of this security ranging strategy depends entirely on the attack detection module, while the random time-hopping ranging module does not increase the complexity at all.

\section{Evaluation}\label{ Simulation Evaluation and Experimental Validation}

\subsection{Numerical Simulation Evaluation}
\label{Numerical Simulation Evaluation}
The signal parameter configurations for simulation are detailed in Table \ref{tab1}.  
The processing flow of the receiver used in this paper includes shaping filtering, downsampling, cross-correlation, timestamp estimation and data demodulation employing a RAKE receiver architecture.
The length of BTW is fixed at 400 samples, with a sampling rate of 2 GHz. 
The MPEP and PAPR thresholds are set to 0.5 and 2, respectively. 
Moreover, the distance between two legitimate devices is 10 m and an attack is considered successful when the measurement result is less than 5 m. 
A total of 20,000 DS-TWR simulations under the \emph{Ghost Peak} attack are performed for each case.
\begin{table}[tbp]
        \caption{Configurations for Legitimate and Attack Signals}
        \label{tab1}
         \vspace{-1mm}
        \small
	\centering
	\begin{tabular}{c|c|c}
        \hline               
			Parameter&Legitimate signals&Attack signals\\      
	\hline              
			Mode &BPRF&BPRF \\
			Preamble spreading factor  &4&9\\
                Preamble code index &	9&	9\\
                SFD number &	0&	0\\
                Modulation	&BPSK+BPM&	BPSK+BPM\\
                Payload encoding & \makecell[c]{RS\\\&Conv. Codes}	&\makecell[c]{RS\\\&Conv. Codes}\\
                Samples of per pulse&	4	&4\\
                Preamble duration&	64	&64\\
                STS segment length	&64&	64\\
	\hline       
        \end{tabular}
\end{table}

SIR is used to denote the ratio between the power of the STS segment of the legitimate signal and that of the attack signal, while SNR indicates the ratio between the power of the UWB signal and noise. 
The simulated attack success probability under different SIR and attack synchronization time $T_{\text{sy}}$ for a typical fixed SNR of -10 dB is shown in Table \ref{tab2}.


\begin{table}[htbp]
        \caption{Simulated Attack Success Probability}
        \label{tab2}
         \vspace{-1mm}
        \small
	\centering
	\begin{tabular}{c cccccc}
        \hline 
          \multirow{2}{*}{$T_{\text{sy}}$} &\multicolumn{6}{c}{SIR (dB)}\\
         \cmidrule{2-7} (\textmu s)& -20&-22&-24&-26&-28&-30\\
	\hline              
	-2.5&4\%&4\%&1.5\%&1\%&1\%&1\% \\
        -2& 4\% &6\% &4.5\% &4\% &1.5\% &1\% \\
        -1.5& 3.5\% &4\% &3.5\% &2\% &2.5\% &1\% \\
        -1& 20.5\% &22.5\% &29\% &32.5\% &21\% &9\% \\
        -0.5& 16\% &22\% &21\% &24\% &23\% & 22.5\% \\
        0& 13\% &14\% &23\% &25.5\% &24.5\% & 19.5\% \\
        0.5& 17\% &23\% &25\% &24.5\% &20.5\% & 19.5\% \\
        1& 11.5\% &20\% &17\% &18.5\% &11.5\% & 6\% \\
        1.5& 5\% &0\% &0.5\% &0\% &0\% & 0\% \\
        2& 4.5\% &0\% &0\% &0\% &0\% & 0\% \\
        2.5 & 2\% &0\% &0\% &0\% &0\% & 0\% \\
	\hline       
        \end{tabular}
\end{table}

\begin{figure}[htbp]              
    \centering   
    \centerline{\includegraphics[width=0.95\linewidth]{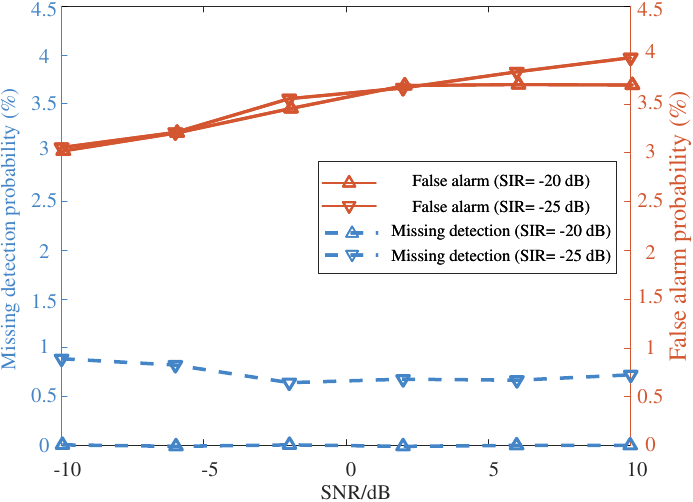}}
    \caption{Missing detection probability and false alarm probability of the attack detection scheme\cite{y15}}
    \label{fign4}
    \vspace{-0.5mm}

\end{figure}

Table \ref{tab2} shows that the attack success probability decreases significantly when the attack synchronization time exceeds a certain range because it affects the detection of the SFD field and the demodulation of the PHR field at the receiver. 
According to the IEEE 802.15.4z protocol, a 128-bit gap exists between the front and the end of the STS segment, which helps attenuate the impact of the attack signal on the SFD field detection and PHR field demodulation. 
Moreover, if the power of the STS segment of the attack signal is too low, the peak value in the cross-correlation spectrum will fall below the thresholds of the leading edge detection algorithm. Conversely, if the power is too high, the strongest path identified by the leading edge detection algorithm may be changed.
Either of these two circumstances will cause attacks to fail.

To evaluate the performance of the proposed attack detection scheme in DS-TWR,  simulations are conducted under the \emph{Ghost Peak} attack condition to obtain the missing detection probability and false alarm probability of the attack detection scheme with respect to SNR. 
As shown in Fig.~\ref{fign4}, the attack detection scheme has a relatively low probability of missing detection and false alarm, and is robust to variations in SNR. 
The reliability of the attack detection scheme provides a solid foundation for applying the random time-hopping DS-TWR.

\begin{figure}[bp]              
    \centering   
    \centerline{\includegraphics[width=0.95\linewidth]{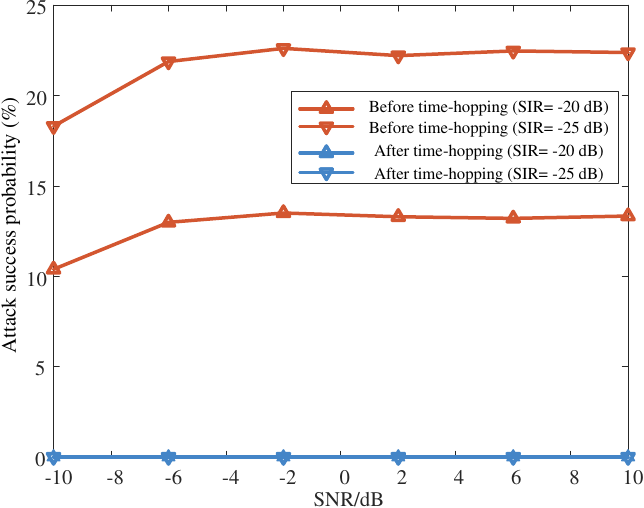}}
    \caption{Attack success probability before and after adopting time-hopping DS-TWR}
    \label{fign6}
    
    \vspace{-2mm}

\end{figure}

The variation curve of attack success probability with SNR before and after using the classical DS-TWR and time-hopping DS-TWR under different SNR and SIR is shown in Fig.~\ref{fign6}, where the random time-hopping delay $ \Delta t$ is set between 15 and 20 \textmu s.  

Fig.~\ref{fign6} indicates that the time-hopping DS-TWR can reduce the attack success probability to 0\% in the simulation involving 20,000 attacks, reaching the order of one in ten thousand regardless of the attack success probability under normal circumstances. This result is consistent with the derived gain $G$. 
The proposed security strategy is robust to variations in both SIR and SNR, verifying that the anti-attack mechanism performs effectively in resisting attacks.


 

\subsection{Experimental Validation}

\begin{figure}[htbp]              
    \centering   
    \centerline{\includegraphics[width=1\linewidth,height=0.9\linewidth]{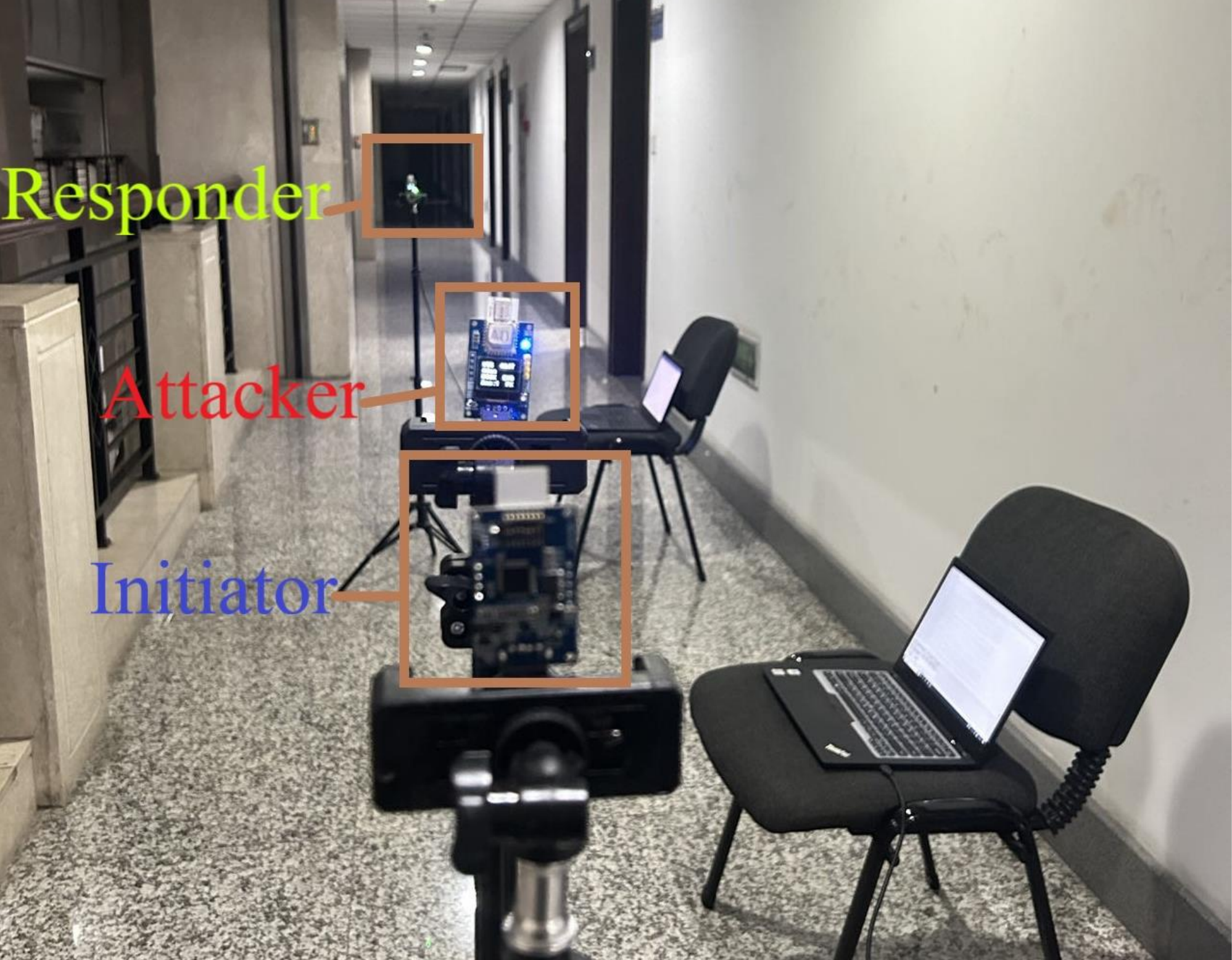}}
    \caption{Practical test environment}
    \label{fig8}
    \vspace{-1mm}
\end{figure}

We validate the proposed security ranging strategy based on random time-hopping mechanism using commercial ranging devices embedded with Qorvo DW3110 chip in a \emph{Ghost Peak} attack scenario conducted in an indoor 10 m Line Of Sight (LOS) environment shown in Fig.~\ref{fig8}.
The attacker is deployed near the Initiator to attack the Response message.  
The results of the attack success probability are shown in Table~\ref{tab3}, with $\Delta t$ set in the range of -2 to 3600 \textmu s in the Mode1 mode and the more vulnerable Super Deterministic Codes (SDC) mode of STS{\cite{g3}}.

According to the listed results, the measurement of the chips can remain stable for tens of hours after adopting the proposed secure ranging strategy without redundant signaling overhead, which indicates that the proposed strategy performs resiliently in the practical environment. 
\begin{table}[htbp]
        \caption{Attack Success Probability of Practical Validation \\(A Total of 10,000 Attacks in Each Case)}
        \label{tab3}
        \vspace{1mm}
	\centering
	\begin{tabular}{c|cc}
        \hline  
         \textbf{STS mode}&\textbf{Befor anti-attack}&\textbf{After anti-attack}\\
         \hline
			 SDC& 60.67\%& 0\% \\
   		Mode1& 0.3\%& 0\% \\
	\hline              

        \end{tabular}
\end{table}

\section{Conclusion}\label{Conclusion}

In order to improve the security performance of existing UWB ranging, this paper proposes the time-hopping DS-TWR which is able to significantly reduce the attack success probability of various types of distance attacks represented by the \emph{Ghost Peak} attack by increasing the randomness of the time of legitimate message transmission during the ranging process. 
Furthermore, combined with the attack detection scheme, the  proposed secure ranging strategy is resilient by adpating to various UWB ranging application scenarios with very low complexity. 
However, this scheme primarily increases the randomness of the transmit time of the ranging message, and future work will incorporate more aspects (e.g. frame format and large ranging networks) to further optimize the security performance of the proposed strategy. 
Considering the current applications of UWB in automotive digital keys, mobile robots, UAV collaborative positioning, and embodied navigation\cite{wg2}, further research on UWB secure ranging methods is expected in the presence of malicious interference or attacks.



\vspace{12pt}

\end{document}